\documentclass[showpacs,preprintnumbers,amsmath,amssymb]{revtex4}
\usepackage{amsmath,amssymb,graphics,epsfig,subfigure}
\usepackage{color}

\begin{document}

\thispagestyle{empty}

\begin{center}

\title{Aschenbach effect and circular orbits in static and spherically symmetric black hole backgrounds}

\date{\today}

\author{Shao-Wen Wei \footnote{weishw@lzu.edu.cn, corresponding author},
        Yu-Xiao Liu \footnote{liuyx@lzu.edu.cn}}

\affiliation{$^{1}$ Key Laboratory of Quantum Theory and Applications of MoE, Lanzhou University, Lanzhou 730000, China\\
$^{2}$Lanzhou Center for Theoretical Physics, Key Laboratory of Theoretical Physics of Gansu Province, School of Physical Science and Technology, Lanzhou University, Lanzhou 730000, People's Republic of China,\\
$^{3}$Institute of Theoretical Physics $\&$ Research Center of Gravitation,
Lanzhou University, Lanzhou 730000, People's Republic of China}

\begin{abstract}
The Aschenbach effect, the increasing behavior of the angular velocity of a timelike circular orbit with its radius coordinate, is found to extensively exist in rapidly spinning black holes to a zero-angular-momentum observer. It also has potential observation in the high-frequency quasi-periodic oscillations of X-ray flux. However, observing such effect remains to be a challenge in static and spherically symmetric black hole backgrounds. In this paper, we mainly focus on such issue. Starting with the geodesics, we analytically study the underlying properties of the timelike circular orbits, and show the conditions under which the Aschenbach effect survives. It is shown that the presence of the static point orbits and stable photon spheres would be the indicator of the Aschenbach effect. We then apply it to three characteristic black holes exhibiting different features. The results state that this effect is absent for both the Schwarzschild and Reissner-Nordstr\"{o}m black holes. While, for the dyonic black hole in quasi-topological electromagnetics, there indeed exists the Aschenbach effect. This provides a first example that such effect exists in a non-spinning black hole background. Moreover, it also uncovers an intriguing property for understanding the black holes in nonlinear electrodynamics.
\end{abstract}


\maketitle
\end{center}

\section{Introduction}
\label{secIntroduction}

Black holes have been a mystery in the study of modern physics since they were predicted by general relativity. They also play a crucial role in understanding observable phenomena, such as shadow, lensing, and Active Galactic Nuclei, in strong gravity region. Meanwhile, via these phenomena, the nature of black holes, as well as the strong gravity, will be potentially tested.

As is extensively known, these phenomena near black holes can be well understood by null or timelike geodesics \cite{Cardoso,Bozza,Hioki}. Amongst them, the circular orbits of test particles are quite interesting. For examples, accretion discs can be appropriately described by stable timelike circular orbits (TCOs), and shadows are related to unstable null circular orbits. Therefore, examining the corresponding properties including the orbital angular velocity, angular momentum, and energy of these circular orbits is quite valuable. In particular, by measuring the angular velocity or angular frequency of the Kepler orbits, the mass and spin of the galactic center black hole SgrA$^*$ can be well determined via the twin high frequency quasi-periodic oscillations \cite{Aschenbach}.

In Newtonian theory, the angular velocity decreases with the radius of a stable circular orbit. This result also holds in the Schwarzschild and Kerr black hole backgrounds for these TCOs observing by a distant observer. However, for a zero-angular-momentum observer or Bardeen observer, it was first noted by Aschenbach that \cite{Aschenbach}, the radial gradient of the angular velocity for the co-rotating TCOs shall present a positive behavior in a narrow radial range when the black hole spin $a/M>0.9953$. Such phenomenon dominated by the rapid spin is now called the Aschenbach effect. Since the required black hole spin is slightly smaller than the Thorne's spin limit $a/M=0.998$ \cite{Thorne}, it has astronomical observation applications. Aschenbach has shown that when the orbital radius is less than one eighth of the gravitational radius, the hump of the radial gradient of the angular velocity approximately coincides with the location, where the radial and vertical epicyclic frequencies have the ratio close to 3:1. Thus, this coincidence could be directly related to the observation of the high-frequency quasi-periodic oscillations of X-ray flux \cite{Stuchlik}.

Soon after the discovery of the Aschenbach effect, it was found in Ref. \cite{Slany} that such effect can also occur for non-geodesic circular orbits with constant specific angular momentum. Such study was also extended to other spinning black hole backgrounds, such as the Kerr-dS/AdS black holes \cite{Mueller} and the braneworld Kerr black holes \cite{Blaschke}. For the charged particles, the Aschenbach effect was found to be enhanced by the negative magnetic field parameter, while weaken by the positive one \cite{Tursunov}. Besides the point particles, such effect was also observed for the spinning particles in the rapidly spinning black hole backgrounds \cite{Khodagholizadeh,Vahedi}.

As shown above, the presence of the Aschenbach effect requires at least two conditions: the black hole must be a rapidly spinning one and the observer should be a zero-angular-momentum observer. This implies that the Aschenbach effect will not appear in a nonspinning counterpart by reducing the black hole spin. On the other hand, recent study exhibits that there exist multi-null circular orbits and more than one pair TCOs for some certain black holes in nonlinear electrodynamics. Such property could provide a chance to test the Aschenbach effect. So in this paper, we aim to investigate such effect to a distant observer in an static and spherically symmetric black hole background. This study can also be a potential test of the universal lower bound on orbital periods around central compact objects \cite{Hod}.

The present paper is organized as follows. In Sec. \ref{coae}, we start with a static and spherically symmetric black hole and obtain the energy, angular momentum, and angular velocity of the TCOs. By making use of these formulas, we show some interesting properties including the stability of these TCOs for a nonspinning black hole. The conditions for the presence of the Aschenbach effect are also studied. Then in Sec. \ref{ssbh}, we take three characteristic black holes as examples, and find that the Aschenbach effect indeed exists in the dyonic black hole background, which is the first observation for a nonspinning black hole. Finally, we summarize and discuss our results in Sec. \ref{Conclusion}.

\section{Circular orbits and Aschenbach effect}
\label{coae}

In this section, we aim to uncover some special properties of the TCOs in a static and spherically symmetric black hole background described by the following line element
\begin{eqnarray}
 ds^2=-f(r)dt^2+f^{-1}(r)dr^2+r^2d\Omega_{2}^{2},\label{metr}
\end{eqnarray}
where the metric function $f(r)$ is only dependent of the radial coordinate $r$. For an asymptotically flat black hole, it behaves as $f(r)\sim 1- 2M/r+\mathcal{O}(1/r^2)$ at infinity with $M$ denoting the black hole mass. The radius $r_{h}$ of the black hole horizon is the largest root of the equation $f(r)$=0. Outside the black hole horizon, we must have $f(r)>0$, while $\partial_{r}f(r)>0$ is not necessary. With the two Killing vectors, $\xi^{\mu}=(\partial_{t})^{\mu}$ and $\psi^{\mu}=(\partial_{\varphi})^{\mu}$, we can obtain two conserved quantities, the energy and orbital angular momentum per unit mass of a massive test particle,
\begin{eqnarray}
 -E&=&g_{\mu\nu}u^{\mu}\xi^{\nu}=g_{tt}\dot{t}, \label{conservedE}\\
 l&=&g_{\mu\nu}u^{\mu}\psi^{\nu}=g_{\varphi\varphi}\dot{\varphi}, \label{conservedL}
\end{eqnarray}
along each geodesic with tangent vector $u^{\mu}$. The dot here denotes the derivative with respect to an affine parameter. Without loss of generality, we need only to concern the equatorial geodesics with $\theta=\pi/2$ for the massive test particle satisfying $g_{\mu\nu}\dot{x}^{\mu}\dot{x}^{\nu}=-1$, which reduces to
\begin{eqnarray}
 -f(r)\dot{t}^2+\frac{1}{f(r)}\dot{r}^2+r^2\dot{\phi}^2=-1, \label{massiveParticles}
\end{eqnarray}
for the metric (\ref{metr}). Combining with above the three equations (\ref{conservedE})-(\ref{massiveParticles}), one has
\begin{eqnarray}
 \dot{t}&=&\frac{E}{f(r)},\\
 \dot{\phi}&=&\frac{l}{r^2},\\
  \dot{r}^2&=&-V_{\text{eff}},
\end{eqnarray}
where the effective potential is given by
\begin{eqnarray}
  V_{\text{eff}} = \left( \frac{l^2}{r^2} + 1 \right) f(r) - E^2.
\end{eqnarray}
At radial infinity, it is easy to obtain
\begin{eqnarray}
  V_{\text{eff}} = 1 - E^2+\mathcal{O}\left(\frac{1}{r}\right).
\end{eqnarray}
Therefore, it is obvious that the energy of a bounded particle is $E\in (0, 1)$ .
For a massive test particle circling the black hole, the angular velocity observed by a distant static observer reads
\begin{eqnarray}
 \Omega=\frac{\dot{\phi}}{\dot{t}}=\frac{f(r)l}{r^2E}.\label{vv}
\end{eqnarray}

\subsection{Time-like circular orbit}
Next, let us focus on the TCO, which requires
\begin{eqnarray}
 V_{\text{eff}}=0,\quad V_{\text{eff}}'=0,\label{vv2}
\end{eqnarray}
where the prime denotes the derivative with respect to $r$. For a given radius of an existing TCO, the corresponding energy and angular momentum can be solved from (\ref{vv2}) \cite{Cardoso}
\begin{eqnarray}
 E&=&\frac{\sqrt{2}f}{\sqrt{2f-rf'}},\label{enan}\\
 l&=&r^{3/2}\sqrt{\frac{f'}{2f-rf'}}.\label{enann}
\end{eqnarray}
Due to the spherically symmetry, the angular momentum of a TCO is symmetric under $l\leftrightarrow-l$, so we only concern the positive one. Moreover, a stable or unstable TCO corresponds to positive or negative $V_{\text{eff}}''$, where
\begin{eqnarray}
 V_{\text{eff}}''=\frac{2(3ff'-2rf'^2+rff'')}{r(2f-rf')}. \label{Vaa}
\end{eqnarray}
The existence of TCOs indicates that the denominator in the above expression (\ref{Vaa}) is positive. As a result, one has
\begin{eqnarray}
 \text{TCO is stable if}  \quad\mathcal{C}>0,\\
 \text{TCO is unstable if} \quad \mathcal{C}<0,
\end{eqnarray}
where $\mathcal{C}=3ff'-2rf'^2+rff''$. Meanwhile, for the marginally stable circular orbit, $V_{\text{eff}}''=0$ gives
\begin{eqnarray}
 \mathcal{C}=0.
\end{eqnarray}
Such particular marginally stable circular orbit denotes the stable circular orbit with the smallest radius that can be continuously connected to spatial infinity by a set of stable TCOs \cite{Delgado}. In general, it coincides with the innermost stable circular orbit (ISCO), while if novel branch of TCO emerges near the black hole horizon, they will be different. In what follows, we would like to address some properties of TCOs by employing Eqs. (\ref{enann}) and (\ref{enan}).

The first property is that the energy and angular momentum diverge at the radius of the photon sphere (PS). Starting with $g_{\mu\nu}\dot{x}^{\mu}\dot{x}^{\nu}=0$ and performing the above similar calculation, one easily finds that the radius of the photon sphere should satisfies $2f-rf'=0$, which indicates that the denominators of $l$ and $E$ vanish. So this property is obvious.

Second, the radius of a TCO is limited in several separated ranges in the radial distance. Positive energy (\ref{enan})  requires $f>0$ and $2f-rf'>0$. The former one is naturally satisfied outside the black hole horizon. Note that $f'<0$ shall lead to imaginary angular momentum for positive energy case. Thus a real $l$ requires both $f'>0$ and $2f-rf'>0$. Combining with them, we find that the radius of a TCO must simultaneously satisfy $f'>0$ and $2f-rf'>0$, which limits the existence of TCOs in several possible separated ranges in $r$.

It was pointed out in Refs. \cite{Weia,Weib} that, for a fixed angular momentum $l$, the stable and unstable TCOs always come in pairs both in rotating and nonrotating black hole backgrounds if they exist. Moreover, the topological argument also suggests that the outermost and innermost ones are radial stable or unstable, respectively. In particular, for each pair TCOs, the stable one and unstable one can be connected or separated. For example, the stable and unstable TCOs for the Schwarzschild black hole is connected by the ISCO. However, if $f'<0$ at such points determined by $V_{\text{eff}}=V_{\text{eff}}'=V_{\text{eff}}''=0$, this pair of TCOs will be separated. Note that different pairs of TCOs are separated as expected from the second property referred above due to the existence of multiple photon spheres.

There is another particular TCO, named as the static point \cite{Weic} with vanished angular momentum or angular velocity. Obviously, it requires $f'=0$, and the corresponding energy is $E=\sqrt{f}$ \cite{Weic,Lehebel}.

One also worths noting that the energy and angular momentum of a TCO are unbounded regardless that the energy is bounded below at radial infinity. Moreover, previous study indicates that the outer pair of stable and unstable TCOs is connected, but the inner pair can be either connected or disconnected. For spinning black holes, we refer to the work in Refs. \cite{Delgado,Lehebel,Collodel}, where the corresponding properties of TCOs and static rings were discussed.

\subsection{Aschenbach effect}

Now, let us turn to the Aschenbach effect. In Ref. \cite{Aschenbach}, Aschenbach found for the first time that, when the Kerr black hole spin is beyond $a/M=0.9953$, the angular velocity of the TCO has a nonmonotonic behavior for a zero-angular-momentum observer or Bardeen observer, while such behavior is absent for a distant observer. Up to now, this Aschenbach effect has not been observed for a nonrotating black hole. In this section, we aim to analyze the conditions of the Aschenbach effect and show that it can exist even for a distant observer in a nonrotating black hole background.

The angular velocity of a TCO for a distant observer can be obtained by inserting the angular momentum (\ref{enan}) and energy (\ref{enann})  into Eq. (\ref{vv}):
\begin{eqnarray}
 \Omega_{CO}=\sqrt{\frac{f'}{2r}}.\label{anve}
\end{eqnarray}
As we have shown above, $f'=0$ is the condition for a static point. If this case occurs, one can see that $\Omega_{CO}$ vanishes at a finite distance. Further considering that it approaches $0^{+}$ at infinity, one can easily conclude that there must exist at least one extremal point of $\Omega_{CO}$, and a branch of TCOs with positive radial gradient of $\Omega_{CO}$ should emerge. A simple algebraic calculation gives
\begin{eqnarray}
 \Omega_{CO}'=\frac{rf''-f'}{2\sqrt{2r^3f'}}.
\end{eqnarray}
So the extremal point is at $rf''-f'=0$. The radial range with monotonically increasing angular velocity should have
\begin{eqnarray}
 rf''-f'>0.\label{ado}
\end{eqnarray}
On the other hand, at some certain values of $r$, the TCOs are absent, so we need to check the TCO can exist even when condition (\ref{ado}) is required. Here, we reformulate the quantity $\mathcal{C}$
\begin{eqnarray}
 \mathcal{C}=2f'(2f-rf')+f(rf''-f').
\end{eqnarray}
If the TCO exists, the first term always positive as mentioned above. Thus the condition (\ref{ado}) leads to $\mathcal{C}>0$. This result states that the Aschenbach effect only occurs in stable TCO branch. While the monotonically decreasing $\Omega_{CO}$ can be both in the stable and unstable TCO branches.

Here we would like to investigate whether the presence of multiple photon spheres is an indicator of the Aschenbach effect. It has been verified that there exists at least one unstable photon sphere or light ring in asymptotically flat spacetime \cite{Cunha2020,Weid}. If multiple photon spheres appears, there must be one stable one. For the convenience of discussion, we define a function
\begin{eqnarray}
 b=2f-rf'.
\end{eqnarray}
Obviously, $b$ vanishes for the photon spheres. Moreover, we have $b'>0$ ($<0$) for the unstable (stable) photon spheres. Let mainly focus on the stable one, the following relation must be hold near it
\begin{eqnarray}
 b'=f'-rf''<0,
\end{eqnarray}
or
\begin{eqnarray}
 rf''-f'>0,
\end{eqnarray}
which is exactly the requirement (\ref{ado}). Therefore, near the stable photon spheres, $\Omega_{CO}'$ could be positive and indicates the existence of the Aschenbach effect.

Another thing worths pointing out is that $\Omega_{CO}$ always maintains finite values at the location of the photon spheres regardless of the infinite energy and angular momentum.

\section{Spherically symmetric black hole }
\label{ssbh}

In the previous section, we have explored the general properties of TCOs, and found that the Aschenbach effect would emerge even for the non-rotating black holes to a distant observer. In what follows, we shall consider several specific black holes, and examine the properties of the angular velocity of TCOs relating with the Aschenbach effect.

\subsection{Schwarzschild black hole}

\begin{figure}
\center{
\includegraphics[width=6cm]{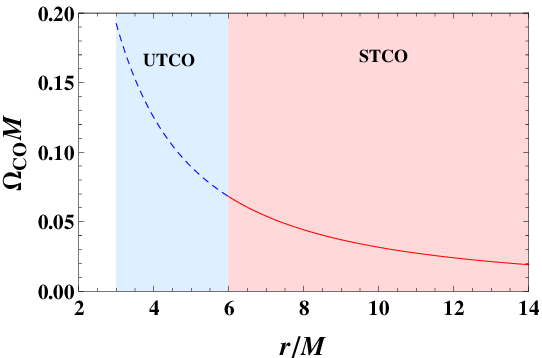}}
\caption{Orbital velocity profile $\Omega_{CO}$ of the TCO in the Schwarzschild black hole background. The characteristic radii are $r_{h}=2M$, $r_{PS}=3M$, and $r_{ISCO}=6M$. The regions in light blue and red colors are for the unstable TCO (UTCO) and stable TCO (STCO) regions, respectively.}\label{pOmeSchBH}
\end{figure}

Since the Schwarzschild black hole is the most known black hole, we now take it as the first example. The line element can be described in the form (\ref{metr}), while with the metric function given by
\begin{eqnarray}
 f=1-\frac{2M}{r},
\end{eqnarray}
where $M$ is the black hole mass. The corresponding effective potential will be immediately obtained
\begin{eqnarray}
 V_{\text{eff}} = \left( \frac{l^2}{r^2} + 1 \right) \left(1-\frac{2M}{r}\right) - E^2.
\end{eqnarray}
For the TCO, we can express its radius in term of the energy or angular momentum. Alternatively, the angular momentum and energy can be expressed as
\begin{eqnarray}
 l&=&\frac{r\sqrt{M}}{\sqrt{r-3M}},\\
 E&=&\frac{r-2M}{\sqrt{r(r-3M)}}.
\end{eqnarray}
In order to guarantee a positive energy, we easily see that $r_{CO}>r_{PS}=3M$, where $r_{PS}$ is the radius of the photon sphere. Furthermore, by solving the condition (\ref{vv2}), one shall find the ISCO orbit locating at $r_{ISCO}=6M$. Thus the radius of the unstable TCOs is in (3$M$, 6$M$), and the stable ones in (6$M$, $\infty$). Obviously, they are connected by the ISCO. On the other hand, the angular velocity of the massive particle along the TCO will be
\begin{eqnarray}
 \Omega_{CO}=\sqrt{\frac{M}{r^{3}}}.
\end{eqnarray}
Obviously, with the increase of $r_{CO}$, $\Omega_{CO}$ decreases. For clarity, we show the angular velocity $\Omega_{CO}$ in Fig. \ref{pOmeSchBH}. The unstable and stable TCO regions are marked in light blue and red colors. The angular velocity $\Omega_{CO}$ in both the regions is a monotonically decreasing function. This indicates that the Aschenbach effect does not exist for the Schwarzschild black hole.

\subsection{Reissner-Nordstr\"{o}m black hole}

When an electric field is included in, one will obtain the charged Reissner-Nordstr\"{o}m (RN) black hole black hole solution by solving the Einstein field equation. The line element can still be described by the form (\ref{metr}), and the metric function reads
\begin{eqnarray}
 f=1-\frac{2M}{r}+\frac{Q^2}{r^2},
\end{eqnarray}
where $M$ and $Q$ are the mass and charge of the black hole, respectively. The event horizon locates at $r_{h}=M+\sqrt{M^2-Q^2}$ obtained by solving $f(r_{h})=0$. It is easy to find when the charge $Q$ tends to $M$, there only leaves an extremal RN black hole. And a naked singularity presents for $Q>M$. Therefore, we require $0\leq Q/M\leq 1$ for the RN black hole.

\begin{figure}
\center{
\includegraphics[width=6cm]{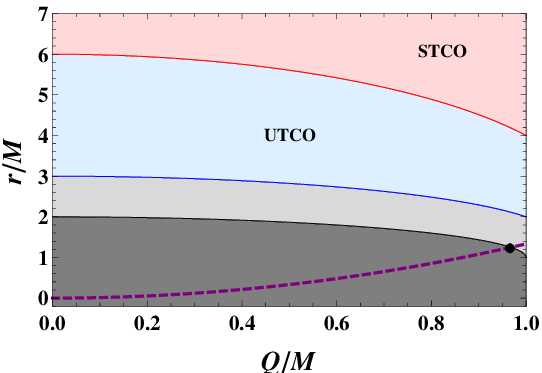}}
\caption{The characteristic radii $r_{h}$ (black thin solid curve), $r_{PS}$ (blue thin solid curve), $r_{ISCO}$ (red thin solid curve) from bottom to top, and $r_{*}$ (purple thick dashed curve) as a function of the black hole charge. The regions in light blue and red colors are for the unstable and stable TCO regions, respectively.}\label{pOmeRN}
\end{figure}

Following (\ref{anve}), the angular velocity is
\begin{eqnarray}
 \Omega_{CO}=\sqrt{\frac{Mr-Q^2}{r^4}}.
\end{eqnarray}
When the charge $Q$ vanishes, the result of the Schwarzschild black hole will be recovered. However, for nonvanishing charge $Q$, $\Omega_{CO}$ has a chance to exhibit a non-monotonic behavior. Such behavior is determined by the extremal point satisfying $\partial_{r}\Omega_{CO}=0$, which gives
\begin{eqnarray}
 r_{*}=\frac{4Q^2}{3M}.
\end{eqnarray}
Therefore, $\Omega_{CO}'$ crosses zero at $r_{*}$, and behaves differently. As the first condition, we require that $r_{*}$ should larger than the radius of the event horizon, i.e., $r_{*}\geq r_{h}$, leading to
\begin{eqnarray}
 \frac{\sqrt{15}}{4}\leq\frac{Q}{M}\leq 1.\label{cadad}
\end{eqnarray}
This can also be observed in Fig. \ref{pOmeRN}, where $r_{*}$ and $r_{h}$ are described in purple and black colors, respectively. For small $Q/M$, $r_{*}$ is below $r_{h}$, while this result reverses when $Q/M$ is beyond $\sqrt{15}/4$. As another requirement, the stable or unstable TCOs should be oustside the photon sphere even when the condition (\ref{cadad}) is satisfied. We also plot the radii $r_{PS}$ and $r_{ISCO}$ in Fig. \ref{pOmeRN}. These TCOs with $r_{PS}<r_{CO}<r_{ISCO}$ (light blue regions) are unstable while these with $r_{ISCO}<r_{CO}<\infty$ (light red regions) are stable. Here, we list the expressions of these two radii
\begin{eqnarray}
 r_{PS}&=&\frac{3M+\sqrt{9M^2-8Q^2}}{2},\\
 r_{ISCO}&=&2M+2\sqrt{4M^2-3Q^2}\cos\left(\frac{1}{3}\arccos\left(\frac{8M^4-9M^2Q^2+2Q^4}{M(4M^2-3Q^2)^{\frac{3}{2}}}\right)\right).
\end{eqnarray}
From Fig. \ref{pOmeRN}, we observe that $r_{*}$ is always smaller than $r_{PS}$, and out of the light blue and red regions. So, the Aschenbach effect will not appear even for a large black hole charge.

\begin{figure}
\center{\subfigure[]{\label{OmeRN09a}
\includegraphics[width=6cm]{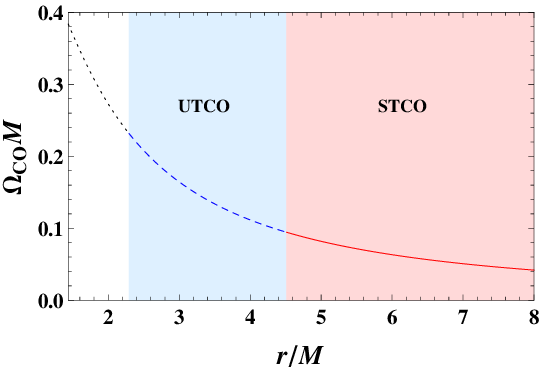}}
\subfigure[]{\label{OmeRN098b}
\includegraphics[width=6cm]{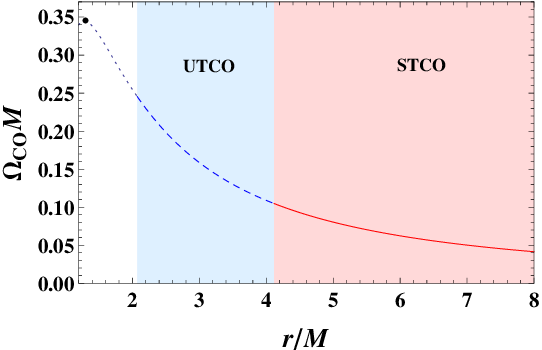}}}
\caption{The orbital velocity profile $\Omega_{CO}$ of TCOs in the RN black hole background. (a) $Q/M$=0.9. The characteristic radii are $r_{h}=1.44M$, $r_{PS}=2.29M$, and $r_{ISCO}=4.51M$. (b) $Q/M$=0.98. The characteristic radii are $r_{h}=1.20M$, $r_{PS}=2.07M$, and $r_{ISCO}=4.12M$. The black dot at $r_*=1.28M$ denotes the local maximum value of $\Omega_{CO}$.}\label{ppOmeRN098b}
\end{figure}

For the sake of clarity, we show the angular velocity in Fig. \ref{ppOmeRN098b} for $Q/M$=0.9 and 0.98, respectively. When $Q/M$=0.9, we find that $\Omega_{CO}$ decreases with $r$. Even in the range ($r_{h}$, $r_{PS}$), where the TCO does not exists, $\Omega_{CO}$ also is a monotonically decreasing function. However, for $Q/M$=0.98, there is a distinct difference. Despite the monotonically decreasing behavior shown in both the stable and unstable TCO regions, there is a monotonic increasing behavior quite near the horizon, ($1.20M$, $1.28M$), which indicates there might exist the Aschenbach effect. Nevertheless, no TCO can survive in this region. Therefore, the Aschenbach effect is still absent. Moreover, a simple algebra shows that, for the extremal black hole with $Q/M=1$, we have $r_{PS}=2M>r_*=4M/3$. This furhter implies that the Aschenbach effect is absent for the RN black hole. Meanwhile, there is only one pair of TCOs, and the stable and unstable ones are connected.

\subsection{Dyonic black hole}

As shown above, both the Schwarzschild and RN black holes do not exhibit the Aschenbach effect. Here as the third characteristic example, we would like to consider the dyonic black hole given in Ref. \cite{Liu2020}. Previous work has shown that there may exist intriguing properties for the motion of particles. There allows the existence of three photon spheres \cite{Liu2020}, and static point orbits \cite{Weic}. This, in some certain, indicates the Aschenbach effect in such black hole background.

To illustrate such result, let us consider the dyonic black hole. The action includes a quasi-topological electromagnetic action term \cite{Liu2020}
\begin{eqnarray}
 S=\frac{1}{16\pi}\int\sqrt{-g}d^4x\left(R-\alpha_1F^2-\alpha_2
   \left((F^2)^2-2F^{(4)}\right)\right),
 \label{action}
\end{eqnarray}
where the field strength is $F^2=F^{\mu\nu}F_{\mu\nu}$ and $F^{(4)}=F^{\mu}_{\;\;\nu}F^{\nu}_{\;\;\rho}F^{\rho}_{\;\;\sigma}F^{\sigma}_{\;\;\mu}$. The coupling parameters $\alpha_1$ and $\alpha_2$ are for the standard Maxwell and quasi-topological electromagnetic actions, respectively. When $\alpha_1=1$ and $\alpha_2=0$, it recovers the standard Maxwell theory. In particular, under the ansatz of global polarization, this quasi-topological term has no influence on the Maxwell equation and the energy-momentum tensor.

The  static spherically symmetric black hole solution to the field equations following from \eqref{action} can also described by the line element (\ref{metr}). The corresponding metric function is given by \cite{Liu2020}
\begin{eqnarray}
 f(r)=1-\frac{2M}{r}+\frac{\alpha_1 p^2}{r^2}+\frac{q^2}{\alpha_1r^2}~_{2}F_1
    \left[\frac{1}{4},1; \frac{5}{4}; -\frac{4p^2\alpha_2}{r^4\alpha_1}
    \right],\label{fr}
\end{eqnarray}
with $~_{2}F_1$   the hypergeometric function. The electric and magnetic charges are $q$ and $p/\alpha_1$, respectively.

\begin{figure}
\center{
\includegraphics[width=6cm]{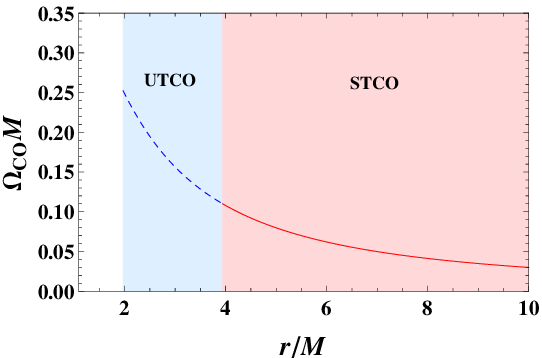}}
\caption{The orbital velocity profile $\Omega_{CO}$ of TCOs in the dyonic black hole background for charge $q/M=1$. The characteristic radii are $r_{h}=1.09M$, $r_{PS}=1.97M$, and $r_{ISCO}=3.94M$. The regions in light blue and red colors are for the unstable and stable TCO regions, respectively.}\label{pOmDyonicBHO}
\end{figure}

Aiming at uncovering the Aschenbach effect, we take $\alpha_1=1,\;\alpha_2/M^2=2.76,\;p/M=0.14$. Further taking $q/M$=1, we plot the angular velocity in Fig. \ref{pOmDyonicBHO}. For this case, we have $r_{h}=1.09M$, $r_{PS}=1.97M$, and $r_{ISCO}=3.94M$. In the range ($r_{h}$, $r_{PS}$), no any TCO can exist. Both in the connected unstable and stable TCO regions, $\Omega_{CO}$ decreases monotonically. Thus the Aschenbach effect is absent for this case.

\begin{figure}
\center{\subfigure[]{\label{OmDyonicBHa}
\includegraphics[width=6cm]{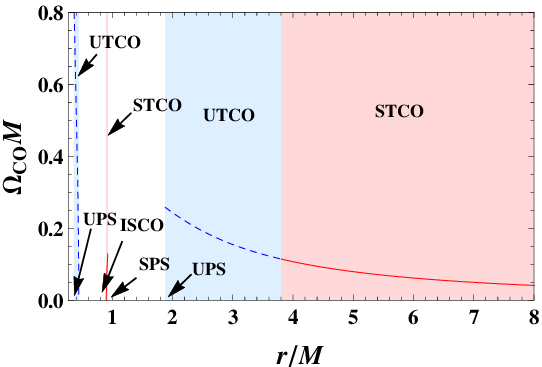}}
\subfigure[]{\label{OmDyonicBHfb}
\includegraphics[width=6cm]{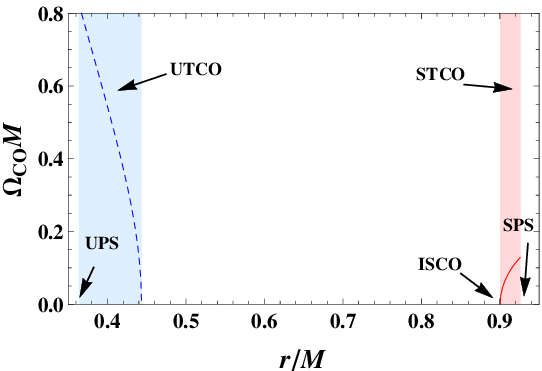}}}
\caption{The orbital velocity profile $\Omega_{CO}$ of TCOs in the dyonic black hole background for charge $q/M=1.02$. The horizon radius is $r_{h}=0.27M$. The regions in light blue and red colors are for the unstable and stable TCO regions, respectively. (a) $\Omega_{CO}$ vs. $r$. (b) Zooming a portion of (a) at small $r/M$. Two unstable photon spheres locate at $r=0.36 M$ and 1.88 $M$, and one stable photon sphere at $r$=0.93$M$. The ISCO is now shifted to $r=0.90M$. ``UPS" and ``SPS" denote the unstable and stable photon spheres.}\label{ppOmDyonicBHfb}
\end{figure}

In Ref. \cite{Weic}, we find that there shall be a pair static point orbits when the black hole charge $q/M\in$ (1.0143, 1.0305). As we shown above, the presence of the static point orbit must be an indictor of the Aschenbach effect. For the purpose, we take $q/M$=1.02 as an example. In previous examples, one easily finds that there exists only one pair unstable and stable regions. However, for this case, beside the outer pair, a novel inner pair near the horizon emerges. These two pairs of TCOs are separated. Moreover, we observe from Fig. \ref{ppOmDyonicBHfb} that the outer stable and unstable TCO regions are connected, but for the narrow inner pair, the stable and unstable TCO regions are separated. The main reason is that there exists a region where these TCOs have complex angular momentum and thus are excluded. Therefore, the stable and unstable TCO regions are separated as expected. This is also a novel phenomenon that not observed before.

We display the angular velocity $\Omega_{CO}$ in Fig. \ref{ppOmDyonicBHfb}. These regions marked in light blue and red colors are for the unstable and stable TCO regions, respectively. The $\Omega_{CO}$ is described with blue dashed curve in unstable TCO regions, and red solid curve in stable TCO regions. In these two unstable TCO regions and outer stable TCO region, $\Omega_{CO}$ decreases, while in the stable TCO region (0.90$M$, 0.93$M$), $\Omega_{CO}$ starts at zero and ends at a finite value. This is clearly an increasing behavior, and thus the Aschenbach effect indeed exists to a distant observer in the static, spherically symmetric black hole background. Furthermore, we have two unstable photon spheres locate at $r=0.36 M$ and 1.88 $M$, and one stable photon sphere at $r$=0.93$M$. The ISCO is at $r=0.90M$.

\begin{figure}
\center{
\includegraphics[width=6cm]{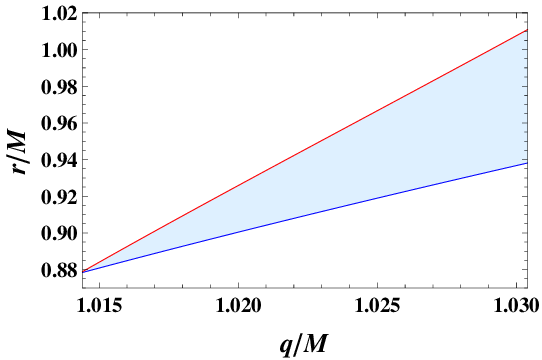}}
\caption{The parameter space of the Aschenbach effect is shown in shadow in the $r$-$q$ plane.}\label{pRqabe}
\end{figure}

In order to examine the effect of the black hole charge on the Aschenbach effect, we show the allowed range of radial coordinate as a function of the charge in Fig. \ref{pRqabe}. When the charge satisfies $q/M \geq 1.0143$, the Aschenbach effect appears. With the increase of the charge, the range of the radial coordinate expands until the maximal charge $q/M=1.0305$ is reached, and beyond which, there only exists the naked singularity.

\section{Discussions and conclusions}
\label{Conclusion}

The Aschenbach effect has been observed for the rapidly rotating black holes to a zero-angular-momentum observer, while not for the non-rotating black holes. In this paper, we aimed to examin it in the the static, spherically symmetric black holes to a distant observer, and discovered for the first time that it indeed exists for the nonspinning dyonic black hole.

First, we considered the circular orbit of a massive particle for a general static, spherically symmetric black hole. With the analytical formula, we demonstrated the relation between the radii of the null and timelike circular orbits. The stability of the TCOs is also analyzed. Moreover, we pointed out that the regions of the TCOs can be connected or separated, which provides us with a novel property of the TCOs. Then we turned to the Aschenbach effect by calculating the angular velocity of these TCOs.  The general result indicates that the Aschenbach effect only occurs in the range of the stable TCO branch. Additionally, if the static point orbits or multiple photon spheres exist, the Aschenbach effect must be observed.

After a general analysis, we took three characteristic black holes as examples.  For the Schwarzschild black hole, the angular velocity decreases with the radius of the TCO. For the RN black hole, the angular velocity admits a non-monotonic behavior. However, the monotonic increasing behavior appears in the region where no TCOs can survive. Thus, there is no the Aschenbach effect for the RN black hole either.

Since the nonlinear electrodynamics attracts much more attention, we considered the dyonic black hole in a theory with a quasi-topological electromagnetic action term. In different parameter ranges, the angular velocity of the TCOs to a distant observer behaves quite differently. For specific example, we set $\alpha_1=1,\;\alpha_2/M^2=2.76,\;p/M=0.14$. If we further took $q/M$=1, the angular velocity is a monotonic increasing function, the same as that for the Schwarzschild and RN black holes. However, by taking $q/M$=1.02, we observed that there are four TCO regions. Two of them admit stable TCOs, while the other two give unstable TCOs. In particular, the outer pair of stable and unstable TCOs is connected, while the inner pair is separated. These two pairs are separated by the existence of the multiple photon spheres. More significantly, in the inner stable TCO region, the angular velocity increases with the radius of these TCOs, obviously indicating the existence of the Aschenbach effect. Furthermore, such effect was also found to be enhanced by the black hole charge.

It should be noted that above results are for the distant observer. One may wonder whether the results still hold for a local Bardeen observer. In order to examine this issue, we need to check the angular velocity $v^{\varphi}=\frac{r}{\sqrt{f}}\Omega_{CO}$ of the circular orbit observed by the Bardeen observer. After a simple calculation, we find that the results remain unchanged. In particular, for the dyonic black hole, the radial regions of the Aschenbach effect for these two observers are the same. However, the angular velocity has an approximately ten times increase for the Bardeen observer. 

Aschenbach effect actually has important influence on the astronomical observations. At the beginning of its discovery, Aschenbach found that it is directly related with the high-frequency quasi-periodic oscillations of X-ray flux \cite{Stuchlik}. On the other hand, we have shown that such effect usually corresponds to the presence of the stable photon sphere, as well as the new regions of the stable TCOs, which could result a novel accretion disk close the black hole horizon. Such light source shall also imprint in observations of gravitational waves and black hole shadow.

In conclusion, we observed the Aschenbach effect to a distant observer in a static and spherically symmetric black hole for the first time. This also uncovers an interesting property for the black hole in the nonlinear electrodynamics. Such effect is also remained to be observed in other black hole solutions in modified gravity, and expected to have potential imprints in modern astronomical observations.

\section*{Acknowledgements}
This work was supported by the National Natural Science Foundation of China (Grants Nos. 12075103, 12105126, and 12247101), and Lanzhou City's scientific research funding subsidy to Lanzhou University.

\end{document}